\documentclass[aps,prl,prabib,twocolumn,showpacs,superscriptaddress]{revtex4}
\usepackage{psfig}
\usepackage{graphicx}
\usepackage{amsmath}
\usepackage{amssymb}
\usepackage{amsfonts}
\usepackage{bm}

\newcommand{\eqname}[1]{\label{eq:#1}}
\newcommand{\bgar}{\begin{eqnarray}}
\newcommand{\enar}[1]{\label{eq:#1}\end{eqnarray}}

\newcommand{\kk}{ {\bf k}}

\newcommand{\xx}{ {\bf x}}
\newcommand{\xixi}{ {\bm \xi}}
\newcommand{\bb} { {\bf b}}

\newcommand{\qq}{ {\bf q}}

\newcommand{\XX}{ {\bf X}}

\newcommand{\eq}[1]{(\ref{eq:#1})}
\newcommand{\al}[1]{^{(#1)}}
\newcommand{\Psihd}{\hat\Psi^\dagger}

\newcommand{\Psih}{\hat\Psi}

\newcommand{\upa}{\uparrow}
\newcommand{\doa}{\downarrow}

\begin{document}

\title{Atom interferometric detection of the pairing order parameter in a Fermi gas}

\affiliation{BEC-INFM, Universit\`a di Trento, 38050 Povo, Italy}
\affiliation{Laboratoire Kastler Brossel, \'Ecole Normale
Sup\'erieure, 24 rue Lhomond, 75005 Paris, France}

\author{Iacopo Carusotto}
\affiliation{BEC-INFM, Universit\`a di Trento, 38050 Povo, Italy}
\email{carusott@science.unitn.it}
\affiliation{Laboratoire Kastler Brossel, \'Ecole Normale
Sup\'erieure, 24 rue Lhomond, 75005 Paris, France}

\author{Yvan Castin}
\affiliation{Laboratoire Kastler Brossel, \'Ecole Normale
Sup\'erieure, 24 rue Lhomond, 75005 Paris, France}

\begin{abstract}
We propose two interferometric schemes to experimentally
detect the onset of pair condensation in a two spin-component Fermi
gas. Two atomic wave-packets are coherently 
extracted from the gas at different
positions and are mixed by a matter-wave beam splitter:
we show that the spatial long range order of the atomic pairs in the
gas then reflects in the atom counting statistics in the output
channels of the beam splitter. 
Alternatively, the same long range order is also shown to create a
matter-wave grating in the overlapping region of the two extracted
wave-packets, grating that can be revealed by a light scattering
experiment. 
\end{abstract}


\pacs{
42.50.-p,  
05.30.Fk,  
39.20.+q,  
74.20.-z,  
}

\date{\today}

\maketitle

The experimental possibility of controlling at will the scattering length $a$
between two spin components of fermionic atoms {\sl via} a Feshbach resonance
has opened the way to a comprehensive study of the
pairing transition in a degenerate Fermi gas \cite{Feshbach,MolecBEC,AtomicBCS}.
The weakly interacting limits are well understood theoretically:
the phase transition
is the Bose-Einstein condensation (BEC) of diatomic molecules ($a=0^+$)
or the BCS transition due to pairing in momentum space ($a=0^-$).
But one can now investigate experimentally the theoretically challenging 
crossover region, including the unitary limit
$|a|=\infty$ ~\cite{CrossoverTheory}.

While the standard techniques used for atomic BECs have allowed to
detect and characterize a molecular BEC~\cite{MolecBEC},
a debate is still in progress about 
experimental signatures of pair condensation
for a negative scattering length.
Several proposals have been put forward \cite{TheoryObsBCS}; none
of them was proved to demonstrate 
the existence of long range order in the pairing parameter.
First experimental evidences of a condensation 
of fermionic pairs in the crossover regime have been recently
presented~\cite{AtomicBCS}, based on a fast ramping of the magnetic 
field to convert pairs on the $a<0$ side into bound molecules on the $a>0$
side, and on the observation of the Bose condensed fraction of the 
resulting gas of dimers. This method is expected to work only when 
the fermionic pairs are small enough, that is in the vicinity of the
Feshbach resonance, $k_F |a|>1$ where $k_F$ is the Fermi momentum.

In this paper, we propose a more direct and general
way of proving the condensation
of pairs, by a measurement of the pairing order parameter, which is 
not restricted to the small pair regime $k_F |a|>1$.
This proposal is the fermionic analog of the atom interferometric
measurement of the first
order coherence function $G^{(1)}$ of a Bose gas \cite{Esslinger}.
More subtle schemes than the observation of the mean atomic density 
have however to be introduced as there is no long range first
order coherence for fermions.
Their experimental implementation would constitute 
a remarkable transposition of
quantum optics techniques to a fermionic matter field.

In the current theories of the superfluid state in fermionic systems \cite{BCS,SupercondBooks,CrossoverTheory},
the onset of pair condensation
is defined by a non-zero long-distance limit $x_{AB}\equiv 
|\xx_A-\xx_B|\rightarrow
+\infty$ of the pair coherence function 
\begin{equation}
  \label{eq:g1pair}
G\al{1}_{\rm pair}(\xx_A,\xx_B)=\left\langle
\Psihd_\uparrow(\xx_A)\,\Psihd_\downarrow(\xx_A)
\Psih_\downarrow(\xx_B)\,\Psih_\uparrow(\xx_B)
\right\rangle,
\end{equation}
this function then factorizing in the product of the order parameter in
$\xx_B$ and the complex conjugate of the order parameter in $\xx_A$.
In the following, we shall propose two distinct methods to measure
$G\al{1}_{\rm pair}(\xx_A,\xx_B)$, both relying on the coherent
extraction of two atomic wave-packets in $\xx_{A,B}$ and their subsequent beating.
The first method is based 
on a two-atom interferometric technique inspired by two-photon
techniques~\cite{Mandel}: it relies on atom counting in the two output channels
of a matter-wave beam splitter.
The second method is based on the coherence properties of light elastically scattered
off the matter-wave interference pattern of the two overlapping wave-packets.

\begin{figure}[htbp]
\begin{center}
\includegraphics[width=6.5cm,clip]{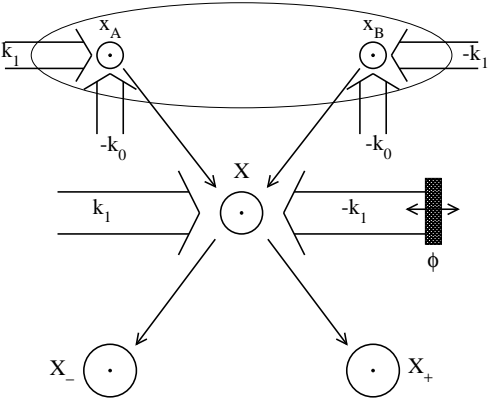}
\caption{First proposed set-up: atoms are
extracted by a Bragg process from the gas at points $\xx_{A,B}$, 
using pairs of laser beams of wavevectors $-\kk_0$, $\kk_1$ and 
$-\kk_0$, $-\kk_1$ respectively; at their overlap position $\XX$,
the two atomic wave-packets are coherently mixed by a laser standing wave 
acting as a 50-50 beam-splitter with adjustable phase shift $\phi$;
the number of atoms in each wave-packet
is measured at the final positions $\XX_\pm$.}
\label{fig:setup}
\end{center}
\end{figure}

Consider a gas of spin-1/2 fermionic atoms at thermal equilibrium in a trap.
At the time $t=0$, the trap potential is suddenly switched off and the
atom-atom interactions brought to a negligible strength, so that the subsequent propagation is
the one of a free atomic field.
At the same time, 
a suitable short pulse of spin-independent optical potential
 is applied (Fig.\ref{fig:setup}) to the atoms situated in regions of size
$\ell_u$ around the points $\xx_A$ and $\xx_B$
so to impart them
a momentum kick of respectively $\kk_0\pm\kk_1$ by means of Bragg
processes and to produce wave packets which are a coherent copy of the field in the trap,
but for a shift in momentum space. $\kk_0$ is taken orthogonal to
$\xx_A-\xx_B$, while $\kk_1$ is parallel to it.
The magnitude $\hbar k_1$ of the counter-propagating momentum kicks
is taken larger than the momentum width $\Delta p$ of the gas,
which is on the order of the Fermi wavevector $k_F$ in the resonance
region ($|a|=+\infty$) and in the weakly interacting BCS regime ($a<0$), 
or on the order of $\hbar/a$ in the case of a molecular condensate ($a>0$).
The size of the extraction
region $\ell_u$ is
taken much smaller than the distance $x_{AB}$ 
between the extraction points. This
latter is taken as macroscopic, that is much larger than any other length
scale of the problem, e.g. the Fermi distance $1/k_F$ and the
Cooper-pair size $\ell_{\rm BCS}$.

In Heisenberg picture the field operator at the end of the optical pulse 
can be related to the initial one by 
\footnote{We assume that $\Delta p |\kk_0\pm\kk_1|\Delta t/m\ll 1$, 
where $\Delta p$ is the initial momentum spread of the gas.}:
\begin{multline}
\label{eq:extracted}
\Psih_\sigma(\xx,\Delta t) =
u(\xx-\xx_A)\,e^{i(\kk_0+\kk_1)\cdot(\xx-\xx_A)} \Psih_\sigma(\xx)+ \\
 +  u(\xx-\xx_B)\,e^{i(\kk_0-\kk_1)\cdot(\xx-\xx_B)}\,\Psih_\sigma(\xx)+\Psih_\sigma^{\rm bg}(\xx).
\end{multline}
The atoms which are left in their original momentum state as well as
the ones having received a different momentum kick during the
extraction process are included in the background field 
$\Psih_\sigma^{\rm bg}$: as
they spatially separate during the evolution, they will be omitted
in the discussion
\footnote{Calculation of expectation values are performed by putting the observables
in normal order and using the fact that $\Psih^{\rm bg}_\sigma$, when acting on the state
vector of the system, gives zero for $\xx$ in one of the two considered wave-packets.
}.
We shall assume for simplicity
that the extraction function
$u(\xixi)$ is a Gaussian, 
$u(\xixi)=u_0\,e^{-\xixi^2/2\ell_u^2}$ of size $\ell_u$; its peak amplitude
$u_0$ is of modulus less than one.

This out-coupling scheme produces two atomic wave-packets traveling with
momentum $\kk_0\pm\kk_1$ starting from respectively
$\xx_{A,B}$. At a time $t_1=mx_{AB}/2\hbar\,|\kk_1|$, the two
wave-packets superimpose around $\XX$.
As mentioned in the introduction, 
the mean density profile in the overlap region does not show fringes so that
more elaborate manipulations have to be performed onto the
atoms in order to measure the pair coherence function $G\al{1}_{\rm
pair}(\xx_A,\xx_B)$.

{\bf Atom-number correlations}: At $t=t_1$, the two overlapping
wave-packets of momentum respectively 
$\kk_0\pm\kk_1$ can be coherently mixed by a spin-insensitive 50-50
matter-wave beam splitter, with reflection and
transmission amplitudes of momentum-independent phase difference $\phi$.
Such a beam splitter may be realized\cite{Bragg}
by applying a pulse of sinusoidal optical potential $U(\xx,t)=4\hbar\Omega(t)\,
\sin^2(\kk_1\cdot\xx+\phi/2)$ 
\footnote{
In order to avoid scattering into higher, non-resonant, momentum
states, the duration $\tau$ of the optical potential pulse has to be
longer than the 
inverse of the atomic recoil frequency $\omega_R$.
In order for the reflection and transmission amplitude to have an equal
modulus and a constant relative phase within the initial momentum
width $\Delta p$ of the gas, $\big|\kk-(\kk_0\pm\kk_1)\big|\lesssim 
\Delta p/\hbar$, $\tau$ has to be short enough for 
$\tau^{-1}\gg \hbar k_1 \Delta p/m$.
The two conditions are compatible since we assumed
that $\Delta p\ll \hbar k_1$.
}.
At a time $t_2$ after the splitting procedure, 
the two emerging wave-packets
of momentum
$\kk_0\pm\kk_1$ will be again spatially separated and centered at 
$\XX_\pm=\XX+\hbar (\kk_0\pm\kk_1)(t_2-t_1)/m$. The total field can be written as
$\Psih_{\sigma}(\xx,t_2) = \Psih^+_{\sigma}(\xx,t_2)+\Psih^-_{\sigma}(\xx,t_2)+\Psih^{\rm bg}(\xx,t_2)$
where the contribution of each packet is
\begin{multline}
\Psih^{\pm}_{\sigma}(\xixi+\XX_\pm,t_2) =\frac{e^{i(\kk_0\pm\kk_1)\xixi}\,e^{i\theta}}{\sqrt{2}}\,
\int\!d\xixi'\,{\mathcal R}(\xixi,\xixi';t_2)\times \\
 \times \,u(\xixi')\,
\Big[\Psih_\sigma(\xx_{A,B}+\xixi')\,+i\,e^{\pm i\phi}\,
\Psih_\sigma(\xx_{B,A}+\xixi')\Big].
\end{multline}
${\mathcal R}(\xixi,\xixi';t)$ is the free-particle propagator and $\theta$
is an irrelevant propagation phase which depends on the details of the
beam splitting procedure. 
The unitarity of ${\mathcal R}$ ensures that the results to come do not depend
on $t_2$.

The operator $\hat{N}_\sigma^\pm$ giving the number of atoms with spin  
$\sigma$ in the wave-packet $\pm$ is obtained by integration
of $\Psihd_{\sigma}(\xx,t_2)\Psih_{\sigma}(\xx,t_2)$ over the spatial
extension of the packet $\pm$ at time $t_2$. 
The operator giving the atom number difference between
the two wave-packets is then $\hat{D}_\sigma=\hat{N}_\sigma^+-\hat{N}_\sigma^-$.
Its expectation value $\big\langle \hat{D}_\sigma\big\rangle$ involves
the first-order coherence function
$\big\langle\Psihd_\sigma(\xx_A+\xixi)\,\Psih_\sigma(\xx_B+\xixi)\big\rangle$ of
the initially trapped atoms,
and therefore vanishes for a macroscopic distance $x_{AB}\gg \hbar/\Delta p$: 
we shall now
take $\langle \hat{N}_\sigma^+\rangle =\langle \hat{N}_\sigma^-\rangle =
N_\sigma$.
Information on the pair coherence function $G_{\rm pair}^{(1)}$
is obtained from the correlation between the two spin components:
\begin{multline}
  \label{eq:up_down}
  C_{\upa\doa}=\langle \hat{D}_\upa\,\hat{D}_\doa\rangle=
\int\!d\xixi\,d\xixi'\,|u(\xixi)|^2\,|u(\xixi')|^2\, \\
\Big[-e^{2i\phi}\,\Big\langle\,\Psihd_\upa(\xx_A+\xixi)\,\Psihd_\doa(\xx_A+\xixi')\,
\Psih_\doa(\xx_B+\xixi')\,\Psih_\upa(\xx_B+\xixi)\,\Big\rangle\\
+\Big\langle\,\Psihd_\upa(\xx_A+\xixi)\,\Psihd_\doa(\xx_B+\xixi')\,
\Psih_\doa(\xx_A+\xixi')\,\Psih_\upa(\xx_B+\xixi)\,\Big\rangle+\textrm{h.c.}\Big].
\end{multline}
From an experimental measurement of the $\phi$ dependence
of $C_{\upa\doa}$, it is therefore
possible to determine whether the system has long range order or not.

An explicit calculation of $C_{\upa\doa}$ as a function of the energy
 gap $\Delta$ can be performed by using the zero temperature
  BCS theory~\cite{BCS,SupercondBooks},  with predictions that are
accurate in the weakly interacting limit only.
In the large $x_{AB}$ limit, only the $\phi$ dependent part of
Eq.(\ref{eq:up_down}) has a non-zero value: $C_{\upa\doa}=C_{\upa\doa}^{(0)}\,\cos(2\phi)$.
In the local density approximation, and assuming
for simplicity that the mean densities
are the same in the two extraction points and in the two spin states,
we find an analytical expression for a wide  
extraction region $\ell_u\gg\ell_{\rm BCS}$, both in the weakly interacting
BCS regime
\begin{equation}
  \label{eq:corr_expl}
 C_{\upa\doa}^{(0)}=
-\frac{3\pi}{8\,\sqrt{2}}\,|u_0|^2\,\frac{\Delta}{E_F}\,N_\sigma
\end{equation}
and in the regime of a molecular condensate
\begin{equation}
C_{\upa\doa}^{(0)} = -\frac{1}{\sqrt{2}}\,|u_0|^2\, N_\sigma.
\end{equation}
$C_{\upa\doa}$ is obtained by averaging over 
many realizations of the whole experimental procedure starting from a trapped gas
in the same initial conditions, so that a knowledge of the signal-to-noise
ratio is relevant.
We estimate the noise 
by the standard deviation of  $\hat{D}_\sigma$:
from Wick's theorem and to leading order in $u_0$,
$\big\langle \hat{D}_\sigma^2 \big\rangle \simeq 2N_\sigma$
which shows that the
shot-noise~\cite{WallsMilburn} in the initial extraction process 
is the dominant source of noise. The number of realizations over which 
to average therefore scales as $(N_\sigma/C_{\uparrow\downarrow}^{(0)})^2$,
which is on the order of $1$ in the BEC limit and on the order of
$(E_F/\Delta)^2$ in the BCS limit.

{\bf Light scattering off the matter-wave grating}:
Information on the pairing coherence function $G\al{1}_{\rm pair}$ of
the trapped gas can also be
obtained by means of light-scattering off the matter-wave interference
pattern formed by the overlapping wave-packets 
at $t=t_1$ around point $\XX$, which is taken in what follows as
the origin of the coordinates.
As already mentioned, the mean density does not show fringes.
On the other hand, fringes appear in the opposite spin density-density
correlation function 
${\mathcal G}\al{2}_{\uparrow\downarrow}(\xx,\xx')=
\big\langle
\Psihd_\upa(\xx,t_1)\,\Psihd_{\doa}(\xx',t_1)
\Psih_{\doa}(\xx',t_1)\,\Psih_{\upa}(\xx,t_1)\big\rangle$. 
As a guideline, one performs an explicit calculation
for BCS theory: one finds that,
as soon as a condensate of pairs is present, fringes
appear as a function of the center of mass coordinates
$(\xx+\xx')/2$ of a pair, with a sinusoidal oscillation of
wavevector $4\kk_1$.
Their amplitude is proportional to the product of the in-trap anomalous
averages in $\xx_A$ and $\xx_B$ and extends up to relative
distances $|\xx-\xx'|$ of the order of the Cooper-pair size
$\ell_{\rm BCS}$.
This matter-wave grating is not easily detected in position space
since its spatial period is smaller than the mean interatomic distance.
We therefore switch to Fourier space:
\begin{equation}
\tilde{{\mathcal G}}\al{2}_{\uparrow\downarrow}(\qq,\qq')
\equiv \int\!d\xx\,d\xx'\,e^{-i(\qq\cdot\xx+\qq'\cdot\xx')}\,{\mathcal
  G}\al{2}_{\upa\doa}(\xx,\xx').
\end{equation}
Since fringes show up on the center of mass coordinate with a wavevector
$4\kk_1$ we limit ourselves to the region $\qq'= \qq\simeq -2\kk_1$
\footnote{More precisely, one assumes
$\hbar qt_1/m \gg \ell_u$ and $\hbar |\qq-2\kk_1| t_1/m \gg \ell_u$.}.
Taking into account the free expansion during $t_1$, one then obtains:
\begin{multline}
\tilde{{\mathcal G}}\al{2}_{\uparrow\downarrow}(\qq,\qq)=
e^{i\hbar\,\Delta q^2 t_1/m}
\int\!d\xixi\,d\xixi'\, e^{-i(\qq+2\kk_1)\cdot(\xixi+\xixi')} \\
\times u^*(\xixi)u^*(\xixi') 
u(\xixi-\bb) u(\xixi'-\bb)\times  \\
\Big\langle\,\Psihd_\upa(\xx_A+\xixi)\,\Psihd_\doa(\xx_A+\xixi')\,
\Psih_\doa(\xx_B+\xixi'-\bb)\,\Psih_\upa(\xx_B+\xixi-\bb)
\Big\rangle
\eqname{G2q}
\end{multline}
where $\bb=\hbar(\qq+2\kk_1)t_1/m$ and $\Delta q=|\qq+2\kk_1|$.
Remarkably, for $\qq=-2\kk_1$ one recovers
the factor in front of $e^{2i\phi}$ in Eq.(\ref{eq:up_down}).

This Fourier component of ${\mathcal G}\al{2}_{\upa\doa}$ 
is detectable in the angular patterns of the elastic light scattering 
from the atomic cloud.
The incoming laser intensity 
has to be weak enough to avoid saturation of the
atomic transition.
Optical pumping processes have to be negligible during the whole
measurement time: the mean number of scattered photons per atom has to
be much less than one not to wash out the information on the internal
atomic state.
The imaging sequence is assumed to take place in a short time
so that the positions of the atoms can be safely considered as fixed.
For each realization of the whole experiment, a different distribution
of the atomic positions is obtained, and consequently a different
angular pattern for the elastic scattering.
Information on the density-density correlation function 
will be obtained by taking the average over many different
realizations.

\begin{figure}[htbp]
\begin{center}
\includegraphics[width=6.5cm,clip]{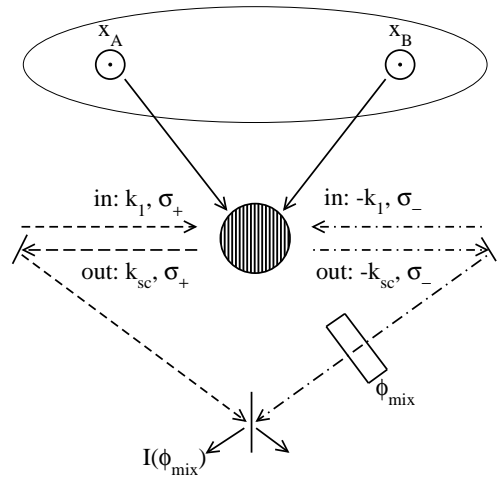}
\caption{Second proposed set-up: 
atoms are extracted from the cloud at $\xx_{A,B}$ and
create a matter-wave grating when the wave-packets overlap at
$\XX$; one detects this grating by shining a pair of
counter-propagating, $\sigma_\pm$ polarized laser beams on the overlap
region and by beating the two 
resulting backscattered light beams on a beam splitter: as function of
the mixing phase $\phi_{\rm mix}$, the beating intensity
$I(\phi_{\rm mix})$  
averaged over many realizations
presents fringes revealing the pair long range order. Dashed (dot dashed) 
lines: $\sigma_+$ ($\sigma_-$) polarization.}
\label{fig:setup2}
\end{center}
\end{figure}

We consider here the simple case when the laser field frequency
is close to resonance with the
transition from the $F_g=1/2$ ground state to 
a $F_e=1/2$ hyperfine component of the excited state.
In this case, $\sigma_\pm$ polarized light interacts only with atoms
respectively in the $\doa,\upa$ spin state. 
The geometry adapted to get information on the condensation of pairs
is shown in
Fig.\ref{fig:setup2}:
a pair of mutually coherent laser beams with a common intensity $I_{\rm inc}$
is sent on the atomic cloud with opposite circular polarizations
$\sigma_{\pm}$ and opposite wavevectors $\pm \kk_1$.
We look at the mutual coherence of two back-scattered beams
in opposite directions $\pm\kk_{\rm sc}$
\footnote{In practice, the scattering wavevectors $\kk_{\rm sc}$ and
$\kk_{\rm sc}'$ can be considered
as opposite if 
$|\kk_{\rm sc}+\kk_{\rm sc}'|<\ell_{\rm BCS}^{-1},
m \ell_{\rm BCS}/\hbar t_1$.}, 
with opposite circular polarizations.
Within the Born approximation (valid if the cloud is optically dilute and
optically thin),
the amplitudes in Fourier space of the scattered light 
on the circular polarizations 
$\sigma_\pm$ are related to the ones of the incoming field by
\begin{equation}
E_{\pm}^{\rm sc}(\pm\kk_{\rm sc})=
A\,{\hat \rho}_{\doa,\upa}(\pm\qq,t_1)\,E_\pm^{\rm inc}(\pm\kk_1),
\end{equation}
where $A$ is a factor depending on the dipole moment of the transition
and on the atom-laser detuning, and
${\hat \rho}_{\sigma}(\qq,t_1)$ is the Fourier component at the transferred
wavevector $\qq=\kk_{\rm sc}-\kk_1$ of the density operator
$\hat{\Psi}^\dagger_\sigma(\xx,t_1)\hat{\Psi}_\sigma(\xx,t_1)$ at time $t_1$.
This mutual coherence is quantified by the correlation function
\begin{multline}
I_{-+}/I_{\rm inc}=\big\langle 
\left[E^{\rm sc}_{-}(-\kk_{\rm sc})\right]^\dagger\,
E^{\rm sc}_{+}(\kk_{\rm sc}) \big\rangle/I_{\rm inc}\\
=|A|^2\,\big\langle
\left[{\hat \rho}_{\upa}(-\qq,t_1)\right]^\dagger
\,
{\hat \rho}_{\doa}(\qq,t_1)
\big\rangle
= |A|^2 
\tilde{{\mathcal G}}\al{2}_{\uparrow\downarrow}(\qq,\qq).
\end{multline}
By using \eq{G2q}, one indeed sees that the correlation function
$I_{-+}$ can reveal the pair long range order.
In the weakly interacting BCS regime as well as in the one of a
molecular condensate, it has a simple
expression for $\qq\simeq -2\kk_1$:
\begin{equation}
I_{-+}=-\frac{|A|^2}{2}
\,C^{(0)}_{\uparrow\downarrow}
\,e^{-\ell_I^2\Delta q^2/2} I_{\rm inc}.
\eqname{Iud}
\end{equation}
As a function of $\Delta q$, it
has a narrow peak with a height
proportional to the correlation function $C_{\uparrow\downarrow}$
of the first proposed set-up and with a width $1/\ell_I$ such that
$\ell_I^2=\ell_u^2+(\hbar t_1/m\ell_u)^2$.
Experimentally, this can be determined
by beating the two scattered beams:  
as a function of the mixing phase $\phi_{\rm mix}$,
the resulting intensity presents oscillations with an amplitude equal 
to $2\,\big|I_{-+}\big|$
on a background of value $\simeq 2 |A|^2 N_\sigma I_{\rm inc}$.

In conclusion, we have proposed two possible ways of detecting
a long-range pairing order in a degenerate Fermi gas by measuring the 
coherence function of the pairs via matter-wave
interferometric techniques.
This has the advantage over other techniques 
of directly measuring the order parameter
without relying on a microscopic description of the many-body state,
so that it applies in an unambiguous way both in the weakly interacting
and the strongly interacting regimes.
More generally, the proposed scheme is an application of quantum optics techniques to
Fermi fields,  a line of research expected to open new possibilities in the experimental
manipulation and characterization of fermionic systems.

\begin{acknowledgments}
We acknowledge helpful discussions with C. Salomon and the members of
his group,
with J. Dalibard, Z. Hadzibabic and L. Carr.
Laboratoire Kastler Brossel is a Unit\'e de
Recherche de l'\'Ecole Normale Sup\'erieure et de l'Universit\'e Paris
6, associ\'ee au CNRS.
\end{acknowledgments}

\end{document}